\documentclass[10pt]{article}
\usepackage{amsfonts}
\usepackage{amsmath}
\usepackage{graphicx}
\usepackage{epsfig}
\usepackage{color}
\usepackage{url}

\begin{document}
	\title{Extremism definitions in opinion dynamics models}
	
	\author{Andr\'e C. R. Martins\\ 
		Universidade de S\~ao Paulo, Escola de Artes Ciências e Humanidades,\\
		Rua Arlindo B\'etio, 1000, Pr\'edio I1, Sala 319-F, 03828-000\\
		S\~ao Paulo, Brazil
	}
	
	\date{amartins@usp.br}
	
	\maketitle
	
	\begin{abstract}
		There are several opinion dynamics models where extremism is defined as part of their characteristics. However, the way extremism is implemented in each model does not correspond to equivalent definitions. While some models focus on one aspect of the problem, others focus on different characteristics. This paper shows how each model only captures part of the problem and how Bayesian inspired opinion models can help put those differences in perspective. That discussion suggests new ways to introduce variables that can represent the problem of extremism better than we do today. 
		
	\end{abstract}
	
	

	\section{Introduction}
	
	Understanding the problems that might arise from strong opinions, such as polarization in political debates \cite{DiMaggio1996,Taber2009,Dreyer2019} or the actions of extremists \cite{tileaga06a,bafumiherron10a} involves both exploring how those questions happen and evolve in the real world as well as obtaining solid and useful theoretical models and definitions. As such, a better comprehension of how opinion dynamics models~\cite{castellanoetal07,galam12a,latane81a,galametal82,galammoscovici91,sznajd00,deffuantetal00,martins08a,martins12b} works and how it can represent those questions is a fundamental step. While we have several models for opinion spread, we need to know how they relate to each other better. Moreover, we also need to define the meaning of key concepts very clearly. Some concepts are easier to define. Consensus simply means agents end up agreeing. Polarization certainly means disagreement \cite{Baldassarri2008}, but it is not as simple as consensus. It also can happen in many different ways \cite{Bramson2017}. In this paper, we will focus on exploring how opinion dynamics models have defined extremist opinions. The consequences and conflicts of those meanings will be discussed.  We will also investigate the need to adopt a definition that is compatible with the meaning of the term outside theoretical models.
	
	Opinion Dynamics models have, since their beginnings, tried to capture how human groups reach consensus or tend to a more polarized state, where disagreement prevails and, often, even gets reinforced. Discrete models~\cite{latane81a,galametal82,galammoscovici91,sznajd00} tend to do a decent job at describing situations where there are two or more clear choices. On the other hand, when we want to know how extreme an opinion is \cite{deffuantetal02a,amblarddeffuant04,galam05,weisbuchetal05,franksetal08a,martins08b,Li2013,Parsegov2017,Amelkin2017}, some kind of continuous variable tends to work better ~\cite{deffuantetal00,hegselmannkrause02}. For situations where we have choices, but we still want to model extreme views, it is common to introduce inflexibles, agents that do not change their opinions \cite{galamjacobs07}. It is also possible to use a model with both discrete and continuous variables that are dependent on one another so that strength and choice can be described \cite{martins08a,martins12b}.

	\section{Different Definitions}
	
	The existence of several competing models poses the question of whether crucial definitions of the relevant variables are compatible. While consensus and polarization are defined in ways that we can easily translate from one model to another, the same is not valid for extremism. Indeed, depending on the mathematical details of the models, it is easier to focus on one aspect of extremism and ignore others. That is what happened.
	
	Typical discrete models are perfect for talking about actions. Polarization is naturally defined  \cite{Bhat_2020}, and the decision to do something or not do it is also binary. As such, they might seem a natural way to model problems such as terrorism. In terrorism, much more important than anything else is the question of whether a terrorist action will happen or not. It seems clear that the strength of opinion plays a fundamental role here. To commit such terrible acts, we expect our agents to be thoroughly convinced about their decisions. However, discrete models are not able to make a statement about the strength of agents' opinions. Instead, in order to capture the essence of extremist opinions, the concept of inflexibles was introduced \cite{galam05}. Here, inflexibility means only that inflexible agents will not change their opinions, regardless of any social influence. The idea does provide a first step towards defining extremism, even if it lacks one crucial aspect.
	
	Continuous models, on the other hand, can naturally speak about opinion strength. If opinions happen on a range of possible values, it is easy to identify the ends of that range as extreme. Of course, unless the range has some practical meaning, such identification might be arbitrary. Nevertheless, given a specific debate and problem, one can often order opinions on that problem over a range in a way that extremes should be near the ends. However, such a definition fails to capture another component of the question.  The kind of extremist behavior we are interested in is often associated with an inability to learn. An agent who has an opinion at the end of the range but who can easily be dissuaded to adopt a central opinion is not a typical extremist. Indeed, continuous models often recognize the fact that extremism does not mean only an opinion at the edge of the spectrum. By changing the range of values over which agents can influence each other or not \cite{Lorenz2010}, extremists can often be associated with smaller uncertainties and, therefore, a smaller capacity to change their minds \cite{Weisbuch2002,deffuant06,martins08c,Ghaderi2014,Sobkowicz2015,Hegselmann2015}.  Mechanisms of stubbornness have also been proposed to address the problem \cite{Ramos2015,Burghardt2016} and allow more realistic extremists. However, all those considerations are usually added in an ad hoc way.

	A central component of the kind of extremism we worry most about, however, is action. Whether we are talking about anti-vaxxers campaigns that threaten to bring back diseases we considered defeated or terrorist attacks, the opinions themselves do not cause any harm. Harm happens when people decide they should act based on their opinions. Opinion models that deal with extremists must be prepared to include questions about conflict, violence, and wars \cite{Morgenstern2013,alizadeh14a}. We need that ability whether the actions we want to describe can be described as terrorism or as something else. Indeed, official sites include several conditions to qualify an action as terrorism \cite{StudyofTerrorism2019}. The need to include actions as central suggests we might need models where we have decision thresholds for extremist actions \cite{Granovetter1978} (not to be confused with thresholds for the number of influencers in an update rule \cite{Watts2007}). As an example, that can be implemented if joining a movement is decided by a utility rule \cite{DeMesquita2005}. Indeed, the action of joining seems central. Important parts of the problem in terrorism can be explained with the use of a  club model, where sacrifices and commitments are required from joiners to ensure non-defection \cite{Berman2008}.  Of course, identifying and modelling networks \cite{Kenney2007,Dixon2010,Cliff2013} including hidden ones \cite{Raab2003} and the echo chambers \cite{Garrett2009,Barbera2015,DelVicario2016,Bastos2017} where extremist discourse happens is fundamental. Nevertheless, proper opinion models can also help us understand strong opinions better. That can help us understand support for all types of ideas and movements, dangerous or not \cite{gargiulomazzoni08a}.
	
	We need definitions that include decisions and strength of opinion in the same model. The possibility of using decision thresholds for joining or supporting extremist groups suggest probability models can be useful in this approach \cite{SIRBU2013}. Indeed, by using probabilities as a measure of opinion and choices based on those probabilities, the Continuous Opinions and Discrete Actions (CODA) model was able to show how extremism can appear even no extremists exist in the initial conditions\cite{martins08a,martins08b}. CODA also shows regular interactions can lead to the appearance of inflexibles  \cite{martinsgalam13a}. More recently, applications of the CODA model to decision problems about actions in health-related problems have been proposed  \cite{Sun2017,Garcia2018}. More than that, the model was also used to propose a formalism that can incorporate any kind of information exchange and agent mental models \cite{martins12b}. As such, it can provide an excellent tool for us to investigate the different meanings extremism has.
	


	\section{Theoretical ways to define extremism}
	
	\subsection{Wishers and Mixers}
	
	Exploring variations of the CODA model can help us understand it better some of the problems with defining extremism. Take, for example, its extension with the introduction of the concepts of wishers and mixers as mental models, coupled with differences in communication \cite{martins16a}. The original CODA model corresponds to wishers, agents who wish to find out which of two options is the best one. The agents assume options are excludent so, if one is best, the other should be avoided. However, if agents were mixers, interested in finding an ideal mix of the two options, we need a different model for how they interpret the opinions of other agents.
	
	Instead of just pointing to the best choice between options $A$ and $B$, mixers assume there is an ideal proportion $p_A$ that should be assigned to $A$. As a consequence,the proportion $p_B=1-p_A$ should be assigned to $B$. What agents communicate in this model is not so clear. They can, in principle, share their average estimate for $p_A$. In this case, with continuous communication and the update of the related continuous variable, we get a Bayesian inspired version of the Bounded Confidence models \cite{martins08c}. However, if we assume agents tell each other which choice they think is the most common one, we still have a discrete choice in the model. In that case, things get more interesting. As the simplest possible mental model, agents might assume that, when observing other agents, they are sampling from a Binomial distribution with probability $p_A$. From their initial opinions on $p_A$, they can use what they observe to update their opinions on $p_A$. 
	
	The simplest way to implement that is to notice that the conjugate distribution to a Binomial likelihood is a Beta $Be(\alpha,\beta)$ distribution with parameters $\alpha$ and $\beta$ associated to   $A$ and $B$, respectively. That means that, if we assume an initial opinion is well described by a Beta distribution, after observing their neighbors, agents will update their opinions and still get a Beta, with just its parameters changed. The update rule becomes quite simple.  When agent $i$ observes $j$ thinks $A$ should happen more often, it adds one to its $\alpha$. If $j$ choice is $B$, one is added to $B$. The most frequent choice is determined by the sign of the difference $\nu=\alpha-\beta$. Therefore, the dynamics of the most frequent choice is obtained by just adding or subtracting 1 to $\nu$, depending on whether $A$ or $B$ is observed as the most common part of the mixture.
	
	The first thing one should notice here is that, aside from an irrelevant normalization parameter, that is exactly the dynamics one finds in the traditional CODA model \cite{martins08a}. The main difference is the meaning of $\nu$, in the original CODA. There, it was defined as $\nu=\ln(\frac{p_A}{1-p_A} )$. However, since the dynamics on both $\nu$s are identical, as well as the choice rule, if one starts with the same initial conditions and use the same random generator draws, the evolution of the choices in the CODA model with wishers will turn out identical to the evolution of preferred choice in the model with mixers. Indeed, after the clusters stabilize, there are also important similarities between both models and the traditional majority model \cite{liggett05,Oliveira1992,galam05}.
	
	However, if one looks at the probabilities one obtains as a result of the two versions, results could not be more different. Translating the model parameters ($\alpha$ and $\beta$, or $\nu$) back to probabilities, one can find that a preferred mix with 93\% of $A$ for mixers can correspond to a probability that $A$ is better around $1-10^{-300}$ for wishers. In both cases, it would take the same amount of interactions for the agents to change their favored state. In the sense of how easy it is for agents in both cases to change their preferences, their opinions are equally extreme. Nevertheless, the comparison of the actual probability values tells a much different story. One mental model leads to practical certainty in favor of an option. The other leads to coexistence, even if the mixing is clearly biased towards one of the choices.

	\subsection{Continuous CODA inspired model}
	
	A similar situation can be observed if we use Bayesian ideas to obtain a model that is similar to the Bounded Confidence class of models \cite{martins08c,Maciel2020}. Here, we can assume agents have an initial opinion represented by a Normal distribution, $N(x_i,\sigma_i^2)$. In order to represent the fact agents in Bounded Confidence model do not trust others whose opinion is too distant, the likelihood includes both a Normal term around the best value $\theta$, that has a probability $p$ associated to it, as well as a Uniform part, that enters with probability $1-p$, representing the possibility $j$ knows nothing about the problem. If we assume agents think other agents are as uncertain as they are, we have
	\begin{equation}\label{eq:likelihooddecept}
	f(x_j|\theta) = p N(\theta,\sigma_{i}^{2}) + (1-p) U(0,1).
	\end{equation}
	From that we can obtain an update rule for the average estimate $x_i$ (assuming only the average and not the whole distribution get updated) given by
	\begin{equation}\label{eq:averagedecept}
	x_i(t+1)=p^*\frac{x_i(t)+x_j(t)}{2}+(1-p^*)x_i(t)
	\end{equation}
	where
	\begin{equation}\label{eq:posteriorp}
	p^* = \frac{p\frac{1}{\sqrt{2\pi}\sigma_i} e^{-\frac{(x_i(t)-x_j(t))^2}{2\sigma_{i}^{2}}} }{p\frac{1}{\sqrt{2\pi}\sigma_i} e^{-\frac{(x_i(t)-x_j(t))^2}{2\sigma_{i}^{2}}} +(1-p)}.
	\end{equation}
	If the distance between the opinions $(x_i(t)-x_j(t)$ becomes too large, $p^*$ tends to zero and $x_i$ basically does not change, representing the lack of trust in distant opinions. If $(x_i(t)-x_j(t)$ tends to zero, on the other hand $p^*$ is basically $p$. For initial trusts close to one, that means the agents will move towards the average value. Those equations lead to a dynamics with the appearance of a number of clustered opinions that depend on $\sigma_i$. Here, $\sigma_i$ plays the role of the threshold factor, representing how easy one agent trust another one with a distant opinion.
	
	What is interesting to observe is that we can also use the Bayesian rule to update $\sigma_i$. We obtain
	\begin{equation}\label{eq:varupdate}
	\sigma_{i}^{2}(t+1)=\sigma_{i}^{2}(t)\left( 1-\frac{p^*}{2} \right) + p^*(1-p^*) \left(\frac{x_i(t)-x_j(t)}{2}\right)^2.
	\end{equation}
	That causes a tendency for $\sigma_i$ to decrease with time. That tendency happens up to the point where all agents become incapable of influencing each other. In other words, agents become the equivalent to agents with very low thresholds. They are unable to update their opinions further. Unlike the ad-hoc models where this kind of stubbornness is limited to the edges of the opinions, where we expect to define extremists, this inability to learn happens for agents with all values of average opinions. They all become inflexibles.

	\section{$M$ possible choices}
	
	In order to discuss the difference between inflexibles and opinions at the extremes, it is also useful to discuss a case where we can model anything from independent choices to choices aligned over an ideological axis. That is possible if we study discrete models with more than two choices. In order to obtain the simplest possible model, the proper choice for variables is the set of $\nu_{q(q+1)}$ defined as the log-odds of the probabilities between two consecutive choices:
	\begin{equation}\label{eq:defnu}
	\nu_{q(q+1)}= \ln{\frac{p(q)}{p(q+1)}},
	\end{equation}
	There are $M-1$ independent parameters $\nu_{q(q+1)}$ and the probabilities $p(q)$ for each choice $q$, where $q=1,\cdots,M$ can be computed back from the $\nu_{q(q+1)}$ by using the set of Equations~\ref{eq:defnu}.
	
	The dynamics that emerge from this model have the same pattern of the traditional CODA model. Local reinforcement causes the appearance of regions where each one of the options is the only supported choice. With the appearance of those regions, opinions inside them tend to become more and more extreme. 
	
	\subsection{Likelihoods and the dynamics of the model}
	
	Equation \ref{eq:defnu} can be obtained from simple considerations on how agents update their opinions on each possible choice $q$ when they observe a neighbor $j$ who prefers choice $m$. While still working with the probabilities of each possible $q$, we can define a likelihood matrix $\mathbb{L}_{mn}=p(a_j=m|x^*=n)$. That is, $\mathbb{L}_{mn}$ represent the mental model each agent has for how other agents' choices are related to the best option or, in other terms, how like $j$ is to prefer $m$ ($a_j=m$), assuming the best choice is $n$ ($x^n=n$).
	
	Using Equation \ref{eq:defnu} to define the consecutive log-odds as basic parameters and applying the Bayes theorem directly to the probabilities $p(q)$, we get an update rule for the opinions of agent $i$ \cite{Martins2020}. When $i$ observes that 
	agent $j$ prefers choice $m$, we have, for every $q=1, \ldots , M$
	\[
	\nu_{q(q+1)}(t+1)=\ln{\frac{p(q,t+1)}{p((q+1),(t+1))}}=\ln{\frac{p(q,t)\mathbb{L}_{mq}}{p((q+1),t)\mathbb{L}_{m(q+1)}}},    
	\]
	and, therefore,
	\begin{equation}\label{eq:bayesmanychoices}
	\nu_{q(q+1)}(t+1)=\nu_{q(q+1)}(t)+\ln \left[ \frac{\mathbb{L}_{mq}}{\mathbb{L}_{m(q+1)}} \right].
	\end{equation}
	
	It is interesting to see that, while the full log-odds likelihood matrix of the interactions, defined as
	\[
	\mathcal{L}_{mqr} = \ln \left[ \frac{\mathbb{L}_{mq}}{\mathbb{L}_{mr}} \right],
	\]
	is what tells us directly how the relation between the probabilities $p(q)$ and $p(r)$ change when the observed agent $j$ prefers $m$, we only need to keep track of the cases where $r=q+1$. That happens because if you know the ratio between $p(1)$ and $p(2)$ and well as the ratio between $p(2)$ and $p(3)$, the ratio between $p(1)$ and $p(3)$ can be trivially obtained and so on. The set of variables $\nu_{q(q+1)}$ is enough to calculate all probabilities.
	
	In the case where the $M$ choices were symmetrical, the symmetry meant that there was only one independent parameter in $\mathcal{L}_{mqr}$ and, as a consequence, in $\mathbb{L}_{mq}$, no matter how large $M$ was. 
	
	However, to see more clearly what the model can say about the definitions of extremism, the symmetrical case is not the most appropriate one. As we will see, if the choices correspond to positions over a one-dimensional ideological line, some questions about what we mean by extreme opinions will appear naturally. Luckily,  the original model with $M$ choices can represent independent choices as well as choices over a one-dimensional ideological line. Intermediate cases or situations that correspond to choices in a higher dimensional ideological space can also be easily introduced. The model for each of those cases is essentially the same. The only difference happens in the values of the likelihoods one uses to define the mental models of the agents.
	
	That is, we need to figure which set of likelihoods $\mathbb{L}_{mq}$  would describe a one-dimensional situation. For example, assume we have $M=5$ possible choices. For the sake of this example, we can identify those positions as extreme-left, left, center, right, and extreme-right, corresponding to the cases $m=1,2,3,4,5$, respectively. Those descriptions might match similar political descriptions, or they might just refer to the positions in a specific drawing about the subject. That suggests that, if one specific value of $m$ is the best choice, it would be natural to expect other agents to pick that choice with higher probability. It is also reasonable to expect that values closer to $m$, such as $m-1$ or $m+1$ should be observed with a larger likelihood than those at a distance of $2$ or more (of course, that is restricted to possible values). Those inequalities assume agents believe others to be more likely to be correct, of course. Situations where one expects others to be extremists and not necessarily correct would lead to a diverse choice of likelihood values and distinct dynamics.
	
	One way to do it is simply to choose by hand values for $\mathbb{L}_{mq}$  that respect those inequalities. However, that means we would have too many free parameters. Arguments of symmetry can diminish the $20$ independent likelihoods in $M=5$ ($25$ conditions $m|q$ minus $5$ requirements probabilities must add to 1) to $9$ or $10$ depending on some assumptions, but that is still too much. In particular, for even larger $M$, the number of possible likelihoods become too large. 
	
	\begin{figure}
		\centering
		\includegraphics[scale=.8]{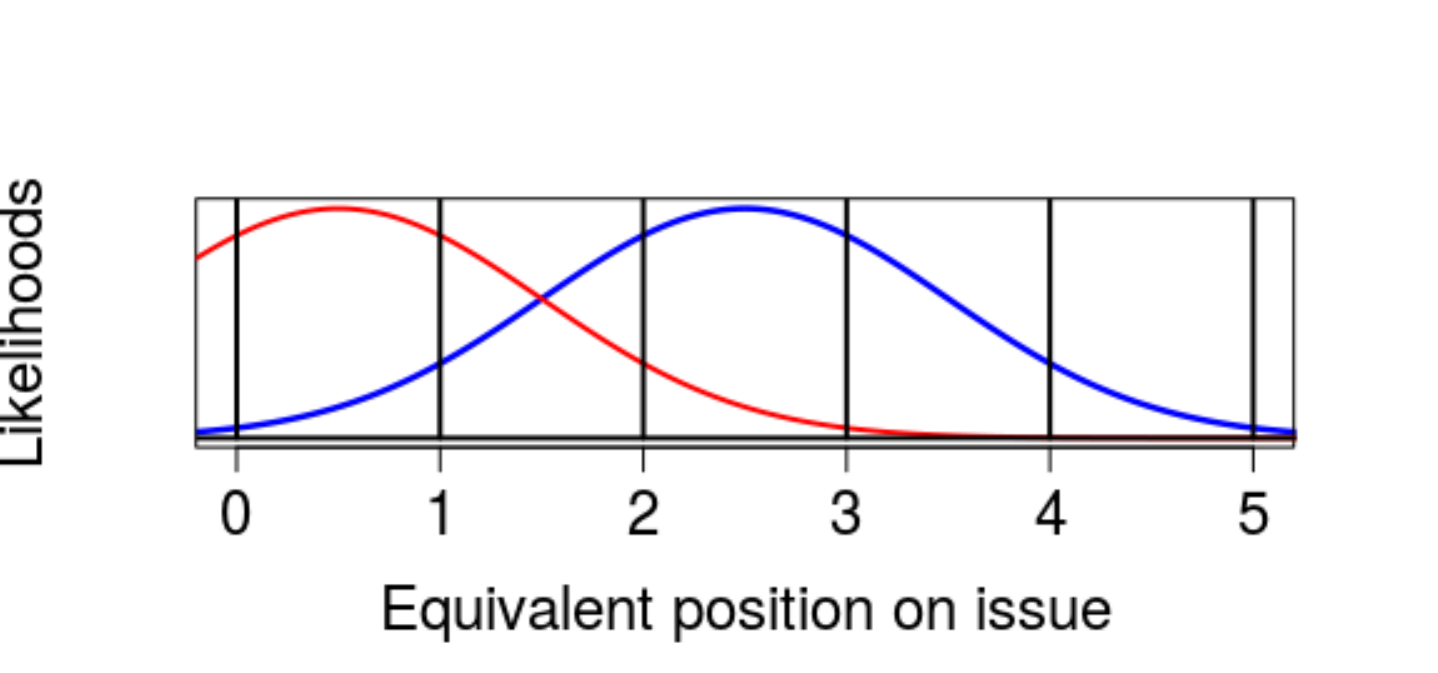}
		\caption{Representation of the one-dimensional choices as consecutive regions over a continuous variable. Possible likelihoods are shown centered on the extreme left choice (red) as well as the central one (blue).
		}\label{fig:likelihooddistr}
	\end{figure}
	
	A much simpler way to model likelihoods that correspond to the expected inequalities is to define a continuous space where the choices correspond to specific ranges and to use continuous probability distributions centered at each choice. Figure \ref{fig:likelihooddistr} shows a representation of that idea, with five regions. In that figure, the first region, which we can call extreme-left, correspond to values over the range $[0,1]$; the left, in the range $[1,2]$, and so, up to the extreme right, in the range $[4,5]$. Possible continuous distributions for this case are shown for the assumptions that extreme-left might be best (red) and that the center might be the best choice (blue).
	
	Likelihoods might be consistently defined simply as proportional to the height of the continuous function at the center of each range or proportional to the area under the distribution (total probability) in the same region. The choice is arbitrary. Both correspond to the case where the inequalities suggested above are respected, as long as a Normal-like distributions are used. That is, any distribution that has only one central maximum, is monotonically decrescent at both sides as we move away from that maximum, and is symmetrical. 
	
	Picking the height of the distribution might be a little easier but there is one question that comes naturally only when we consider the actual probabilities (areas). There are parts of the distributions that fall outside the range of all $M$ boxes. One can, of course, use only the boxes and renormalize. But, in circumstances where there is an expectation that extreme opinions might be more common than they should be, including the tail probabilities in the extreme cases might be used to represent that. That possibility can generate cases where our desired inequalities are not satisfied, however. Both cases, renormalized and with tail probabilities included in the extreme values, were included in the initial exploratory simulations. There were no qualitative differences, however. Tail probabilities just exacerbated the observed behavior for large likelihood uncertainty. That behavior will be described below.
	
	The first thing one learns when trying to use unbounded extreme boxes is the importance of the difference in likelihoods. Assume each box has a size of one and that the chance each choice is observed is given by a Normal or a t-Student distibution centered on the box with a standard deviation given by $\sigma_l$. As initial exploratory simulations showed that the degrees of freedom of a t-distribution had a similar effect to the standard deviation, all cases in this article correspond to a t-distribution with 30 degrees of freedom, which is reasonably close to a Normal distribution. 
	
\begin{figure}
		\centering
		\begin{tabular}{ccc}
			\includegraphics[scale=.18]{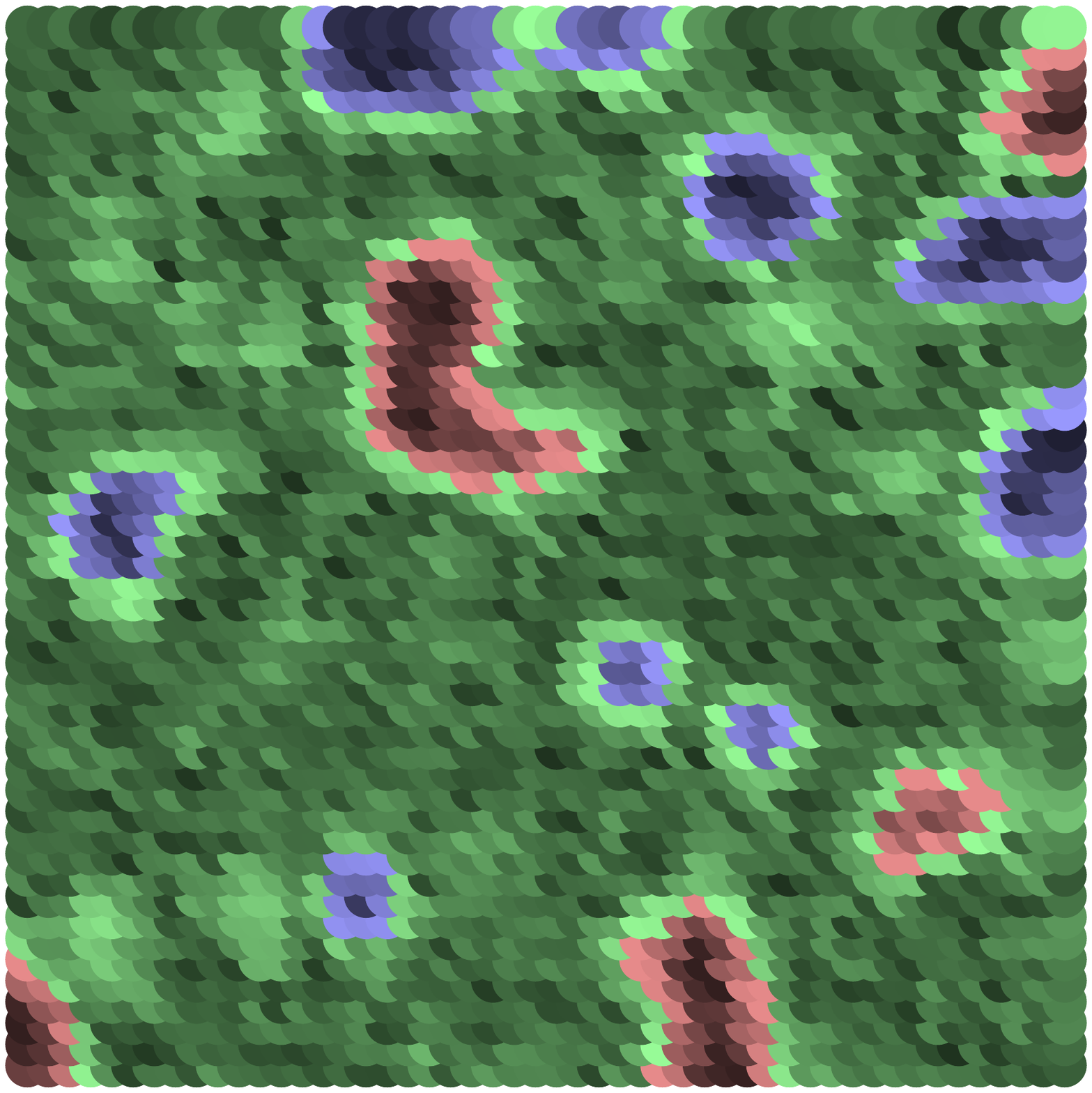}&
			\includegraphics[scale=.18]{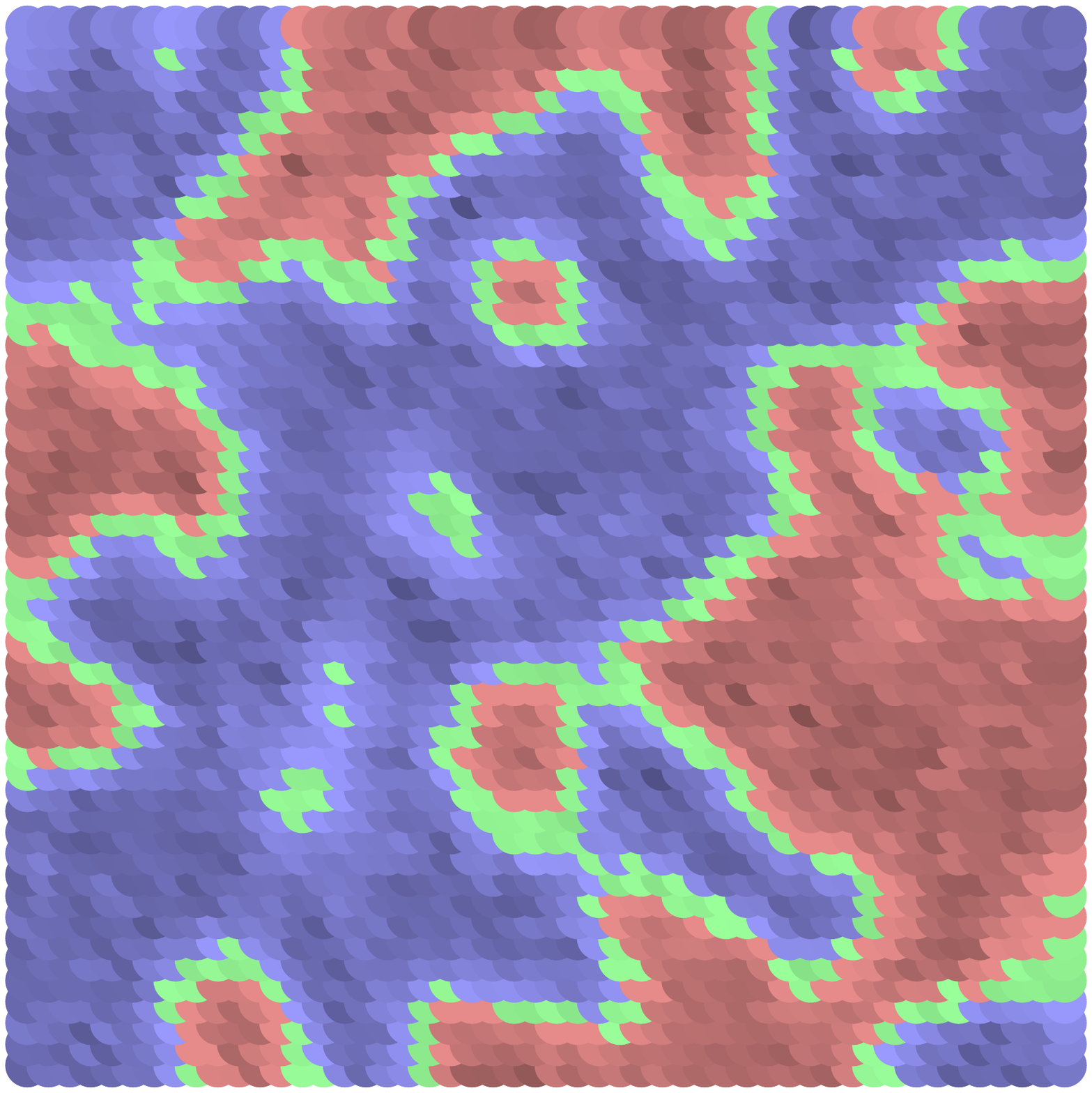}&
			\includegraphics[scale=.18]{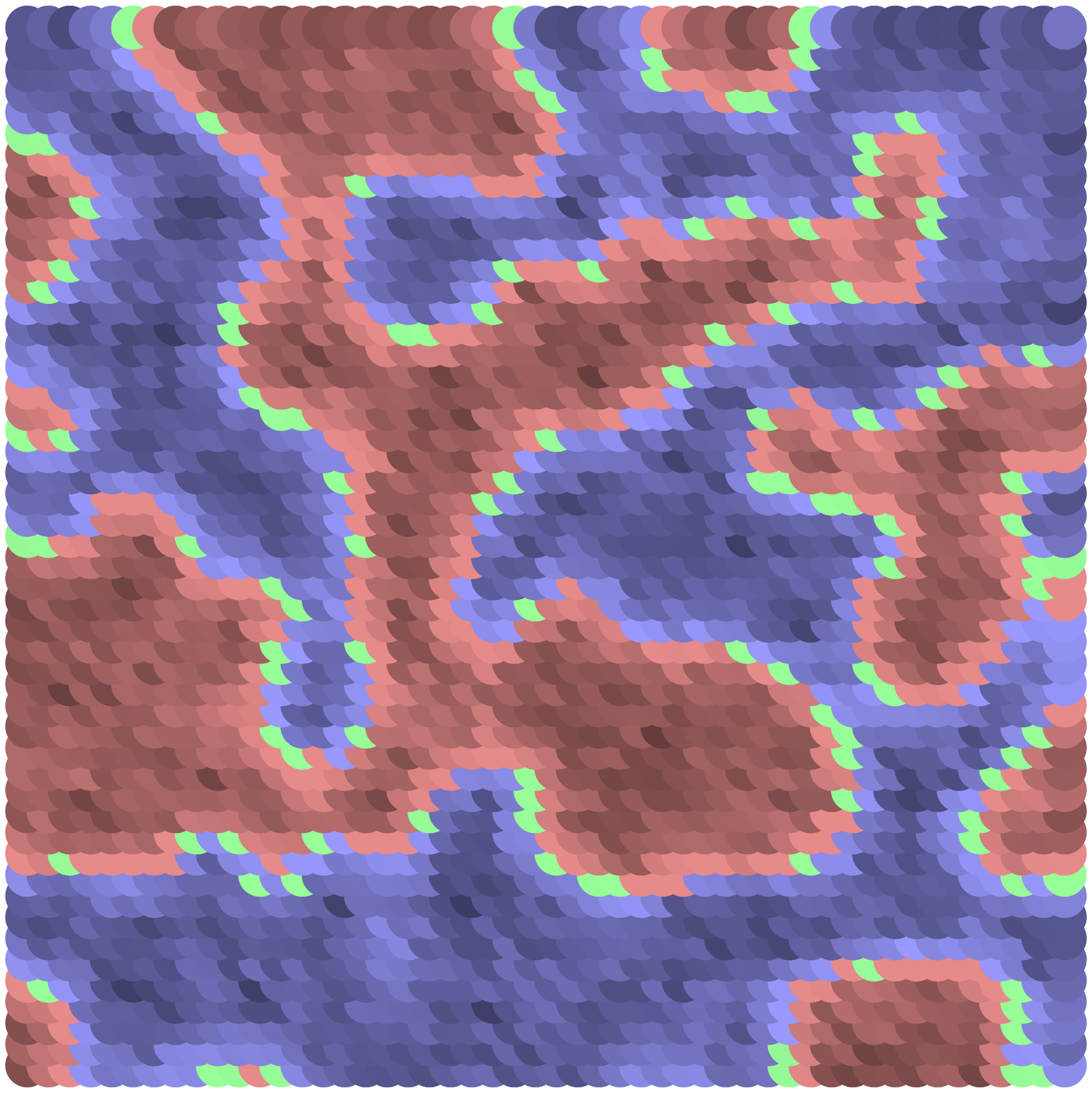}\\
			\includegraphics[scale=.18]{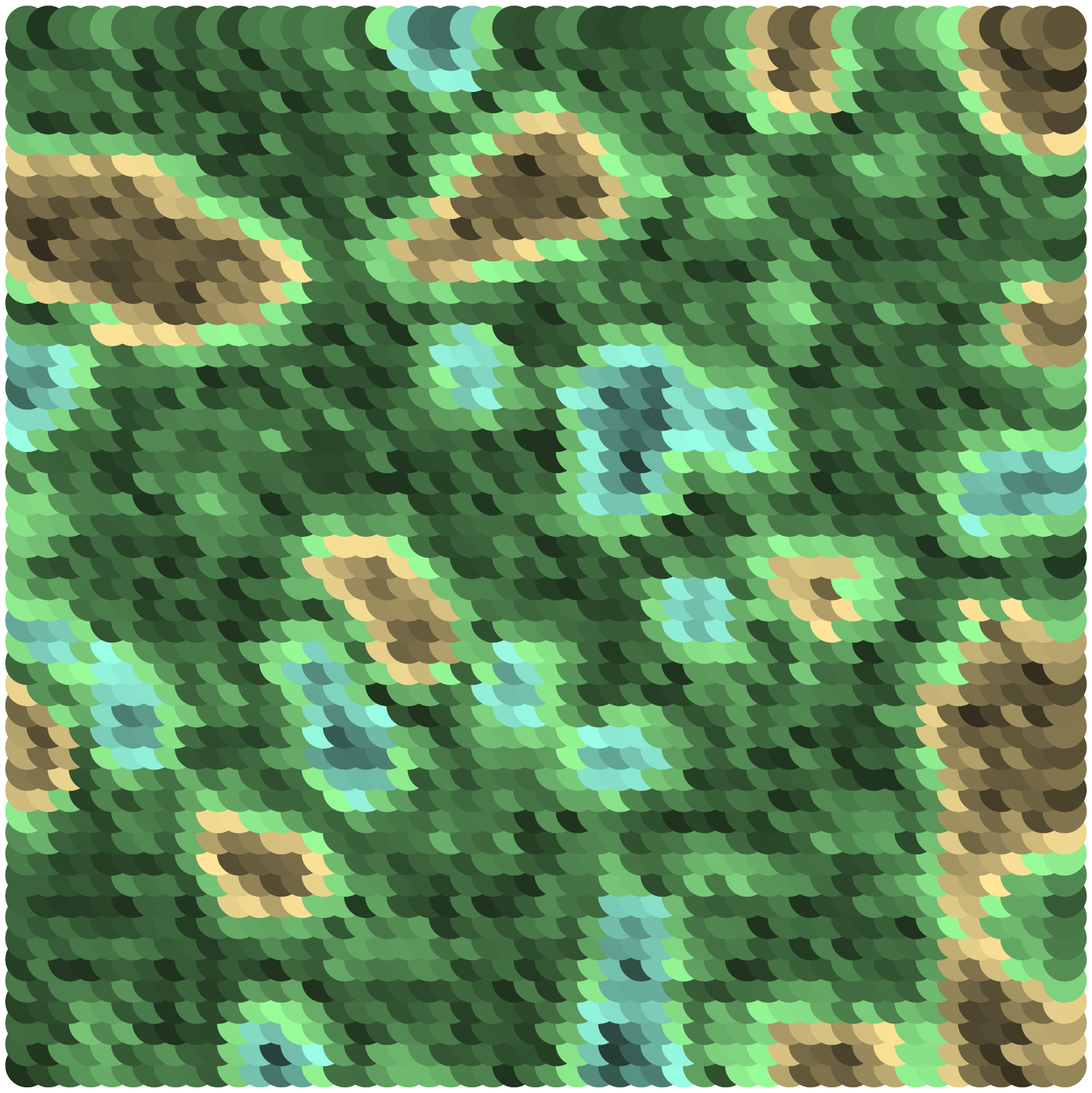}&
			\includegraphics[scale=.18]{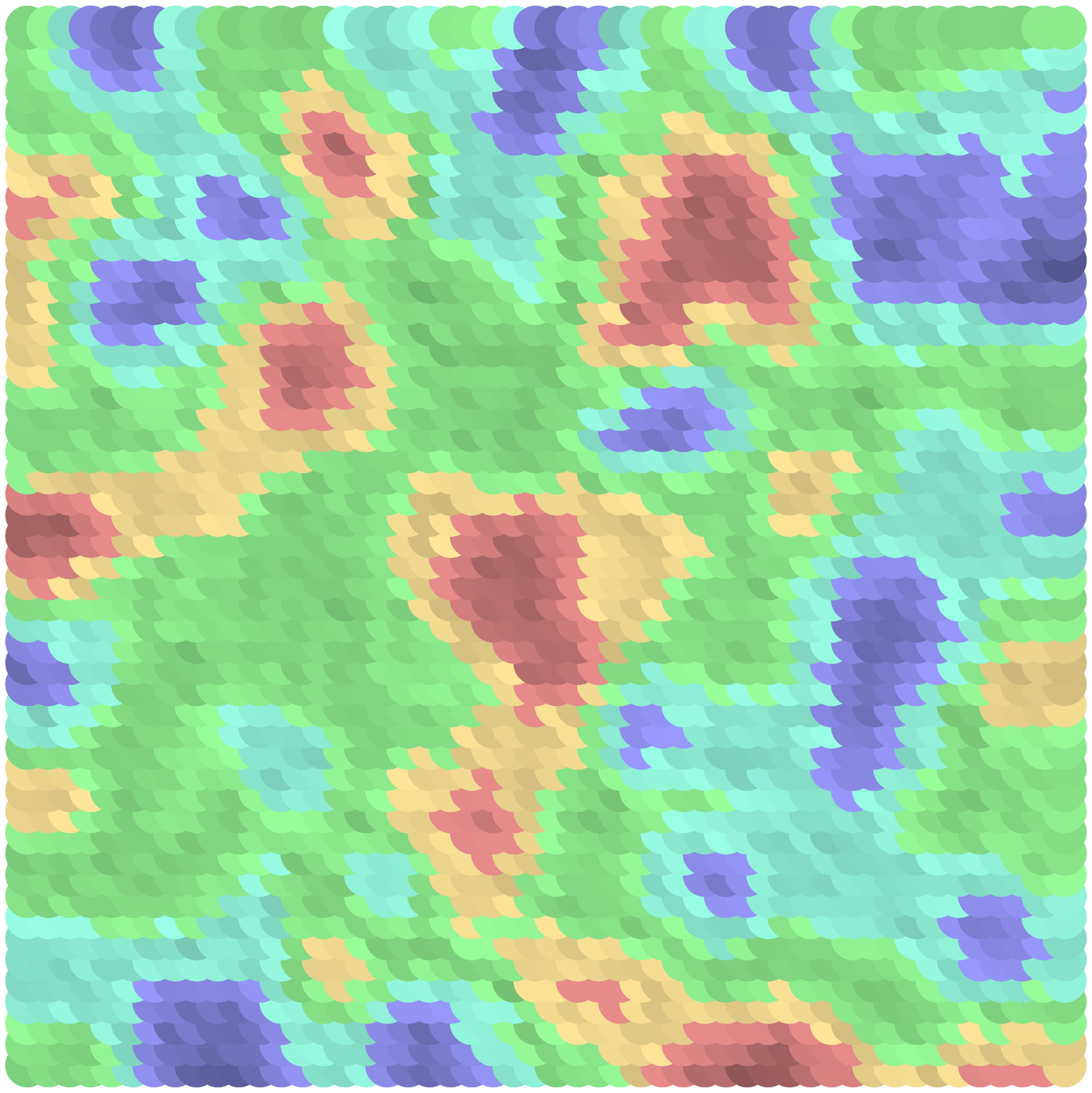}&
			\includegraphics[scale=.18]{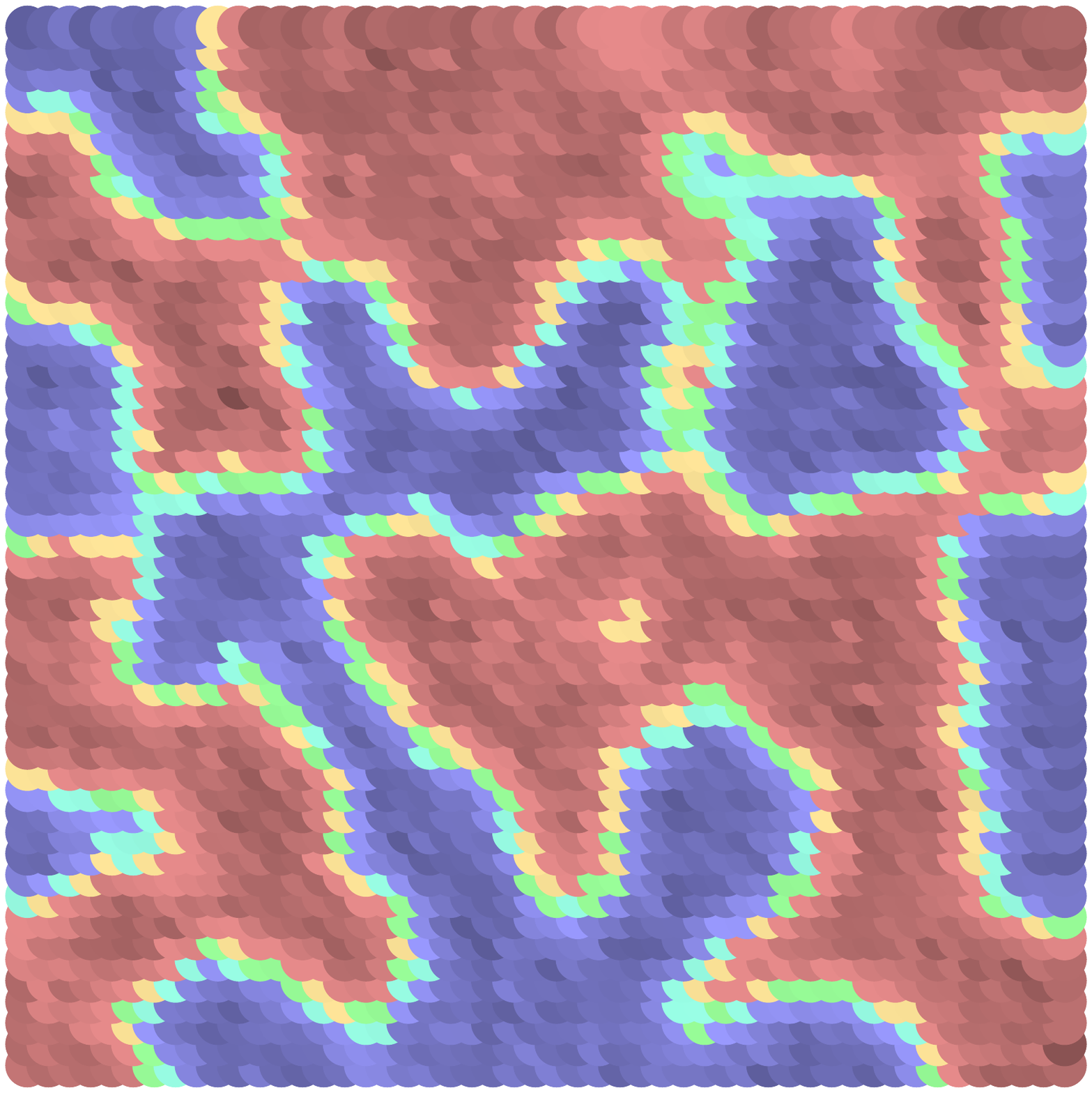}
		\end{tabular}
		\caption{Opinion landscapes showing final preferences for specific realizations of the $M$ choices model after $T=50$ average interactions per agent. Each color corresponds to a specific choice, while opinion strength is shown by a lighter (weaker opinion) to darker (stronger) hue of the collor. Left column figures correspond to $\sigma_l=0.5$, central columns, to $\sigma_l=1.0$ and right columns to $\sigma_l=2.0$. {\it Top line}: Case where $M=3$ is shown, colored in tones of red, green, and blue. {\it Bottom line}: Case where $M=5$ is shown, colored in tones of red, yelow, green, cyan, and blue.
		}\label{fig:oplandscapes}
	\end{figure}

	It is easy to observe in all implementations that, for large values of $\sigma_l$, the central opinions tend to almost disappear in the long run. Central opinions survive only in transition regions between the more extreme options. That behavior can be seen in Figure~\ref{fig:oplandscapes}, where the final preferences for specific realizations of the $M$ choices model after $T=50$ average interactions per agent are shown. The first line corresponds to $M=3$ cases, and the second one to $M=5$. When $M=3$, preferences correspond to the ordered colors red, green, and blue.  When $M=5$, the colors representing each choice were red, yelow, green, cyan, and blue, ordered from extreme left to extreme right. Strength of opinion for choice $m$ was measured as the log-odds of $m$ against all other possibilities, that is
	\[
	\nu_{m}= \ln{\frac{p(m)}{1-p(m)}}.
	\]
	Darker hues correspond to stronger opinions, that is, larger values of $\nu_m$. Correspondence between hues and values of $\nu_m$ are specific to each landscape, chosen for ease of visualization.
	Different likelihoods correspond to distinct columns. The left column shows the cases where  $\sigma_l=0.5$; central columns, $\sigma_l=1.0$; and right columns, $\sigma_l=2.0$. While the figures correspond to only one realization, the same behavior was observed in each case over 20 repetitions of each parametrization.
	
	The most apparent feature of those landscapes is the fact that, while for $\sigma_l=0.5$, the most common surviving choice is the central green color choice ($m=2$ when $M=3$ and $m=3$ when $M=5$), that tendency gets reversed as  $\sigma_l$ increases. For $M=3$, a value of $\sigma_l=1.0$ is enough to make the green choice just a boundary between the red and blue choices and as $\sigma_l=2.0$,  even a good part of the green boundary disappears. Almost no centrists seem to be left. For $M=5$, we see an intermediate state when $\sigma_l=1.0$, where all colors seem to coexist. When $\sigma_l=2.0$, however, we observe once more that we only have regions of the extreme choices, red and blue. The three intermediate options survive only in the boundaries between red and blue, the same as in the $M=3$ case.

	To understand why that happens, one must look at the likelihood values in Table~\ref{table:likelihoods} and the log-likelihoods one obtain from those. We can see that, when we have $\sigma_l = 1.0$ and an agent who supports $m=1$ is observed, that causes a change in the variables so that $\nu_{12}$ (evidence supporting 2 over 1) decreases by approximatedly 0.69 and $\nu_{23}$ decreases by 1.1. Of course, $\nu_{13}$ changes by decreasing 1.79, but that is a consequence of the two other values. On the other hand, when agent $j$ supports $m=2$, we have $\nu_{12}=+0.24$ and, symmetrically $\nu_{23}=-0.24$.  When agent $j$ supports $m=3$, we have the symmetrical, opposite values for when $m=1$.
	Similarly, when  $\sigma_l = 0.5$, we have that, when an agent who supports $m=1$ is observed, $\nu_{12}=-1.6$ and $\nu_{23}=-3.9$. If agent $j$ supports $m=2$, however, $\nu_{12}=+1.3$ and, symmetrically $\nu_{23}=-1.3$.

	\begin{table}[ht]
		\caption{Likelihoods for different values of  $\sigma_l$} 
		\centering 
		\begin{tabular}{c c c c} 
			\hline\hline 
			& $\sigma_l = 0.5$  &  &\\
			Likelihoods & 1 & 2 & 3 \\ [0.5ex] 
			\hline 
			1 & 80.6\% & 16.1\% & 0.3\% \\ 
			2 & 19.1\% & 67.8\% & 19.1\% \\
			3 & 0.3\% & 16.1\% & 80.6\% \\[1ex] 
			\hline\hline 
			& $\sigma_l = 1.0$  &  &\\
			Likelihoods & 1 & 2 & 3 \\ [0.5ex] 
			\hline 
			1 & 55.7\% & 27.8\% & 9.3\% \\ 
			2 & 35\% & 44.3\% & 35\% \\
			3 & 9.3\% & 27.8\% & 55.7\% \\[1ex] 

			\hline 
		\end{tabular}
		\label{table:likelihoods} 
	\end{table}
	
	That is, when we pick a small $\sigma_l$, it is expected that, when choice $m$ is correct, it will be observed more often than it would be the case for a larger $\sigma_l$. That has a particularly significant effect on central choices. As $\sigma_l$ gets larger, extreme opinions are still distant from one another and that means it is easier to distinguish them by observing the choices of the neighbors. The same is not valid for central choices. Notice that when $\sigma_l = 1.0$, observing an extreme choice drives the log-odd between the extreme and central opinions of agent $i$ by 0.69. If the central choice is observed, however, the log-odds in favor of the central opinion, however, increases only by 0.24. That means observing an extremist has a much stronger influence than a centrist (a factor of almost three). For $\sigma_l = 0.5$, there is still a difference, but much weaker (1.6 versus 1.3). In that case, the initial random configuration, where most agents start in a mixed neighborhood, seems to lead the system to a state where the central opinion is preferred. The same analysis can be performed when $M=5$.
	
	While mapping the cases where each opinion becomes predominant is a fascinating subject in itself, for this paper, the case where $\sigma_l = 0.5$ is the most illustrative one. We can see in the first column of Figure~\ref{fig:oplandscapes} that not only the central green choice dominates, it also creates regions where the agents reinforce their opinions in favor of the central choice. While lighter shades of green are observed in the boundaries, agents who only observe centrists do gain a firm opinion in favor of their choices. Given enough interactions, they can easily be labeled as extremists in favor of the central choice. Indeed, it might be difficult to change their choices. Those agents do become inflexibles in a practical sense, just as inflexibility can emerge when in the simpler $M=2$ case~\cite{martinsgalam13a}. If those opinions in favor of the central choice are extreme or not depends on how one defines the term.

	\section{Discussion}
	
	Opinion dynamics needs a better definition of extremism. Our models do not agree with each other on what it means. However, the reason for the disagreement is not that the definitions in some of those models are wrong. Instead, they capture different aspects of the same problem. Extremists rarely change their positions. Models can model that either as actual inflexibles  \cite{galamjacobs07} or as opinions so strong they behave as if they were inflexibles \cite{martinsgalam13a}.
	On the other hand, when we talk about the ideological axis, extremism is indeed usually strongly associated with opinions near the edges. That is not a necessity, however, and it is possible to have extremist behaviors defending all kinds of values or choices over that axis, even central ones. That is true, at least in principle. Indeed,  in the model with $M$ choices, those choices can be independent. In that case, there would be no actual central choice, and we would quickly agree that extremism can happen in favor of all preferences. From a modeling point of view, the only difference between independent choices and choices aligned on one ideological axis (or, possibly, on a multiple dimensional ideological space) rests on the choices of likelihoods. An agent can become extremist in favor of the central choice.
	
	There is one more aspect we must introduce in our models. Having strong and stubborn opinions is not enough to make someone an extremist. Being certain is not the same as being extreme. There are problems where stubborn opinions are not a problem at all. They just mean agents might be very sure about something. When certainty is warranted, strong opinions might even be correct. However, they do become problematic when associated with either very distorted views of the world or with the decision to commit violent actions. Violence or not, problematic extreme views are always linked to acting on those views. So, we need frameworks that allow us to speak of the strength of opinions, choices, and actions. Actions are a particular type of choice that can be easily implemented in a Bayesian inspired model. That is why the problems with definitions became much clearer when we looked at extensions of the CODA model. As the meaning of terms become better defined and implemented, we should be able to propose better models for human exchanges of information. One first fundamental step towards more realistic models is to define extremism in a way that captures its many distinct characteristics.


	\bibliographystyle{unsrt}
	\bibliography{biblio}

\end{document}